\documentclass{article}

    \usepackage[final]{mlsafety_neurips_2022}

\usepackage[utf8]{inputenc} % allow utf-8 input
\usepackage[T1]{fontenc}    % use 8-bit T1 fonts
\usepackage{hyperref}       % hyperlinks
\usepackage{url}            % simple URL typesetting
\usepackage{booktabs}       % professional-quality tables
\usepackage{amsfonts}       % blackboard math symbols
\usepackage{nicefrac}       % compact symbols for 1/2, etc.
\usepackage{microtype}      % microtypography
\usepackage{xcolor}         % colors
\usepackage{graphicx}

\title{Adversarial Attacks on Transformers-Based Malware Detectors}

\author{%
  Yash Jakhotiya\thanks{Corresponding author}\\
  Department of Computer Engineering,\\
  College of Engineering, Pune\\
  \texttt{jakhotiyays16.comp@coep.ac.in} \\
   \And
   Heramb Patil \\
   Department of Computer Engineering,\\
   College of Engineering, Pune \\
   \texttt{herambnp16.comp@coep.ac.in} \\
   \And
   Jugal Rawlani \\
   Department of Computer Engineering,\\
   College of Engineering, Pune \\
   \texttt{rawlanijr16.comp@coep.ac.in} \\
   \And
   Dr. Sunil B. Mane \\
   Department of Computer Engineering,\\
   College of Engineering, Pune \\
   \texttt{sunilbmane.comp@coep.ac.in} \\
}

\begin{document}

\maketitle

\begin{abstract}
Signature-based malware detectors have proven to be insufficient as even a small change in malignant executable code can bypass these signature-based detectors. Many machine learning-based models have been proposed to efficiently detect a wide variety of malware. Many of these models are found to be susceptible to adversarial attacks - attacks that work by generating intentionally designed inputs that can force these models to misclassify. Our work aims to explore vulnerabilities in the current state of the art malware detectors to adversarial attacks. We train a Transformers-based malware detector, carry out adversarial attacks resulting in a misclassification rate of 23.9\% and propose defenses that reduce this misclassification rate to half. An implementation of our work can be found at 
\url{https://github.com/yashjakhotiya/Adversarial-Attacks-On-Transformers}. 
% the anonymized repository  \url{https://github.com/anonauthor/Adversarial-Attacks-On-Transformers}.
\end{abstract}

\section{Introduction}

Malware is software written to steal credentials of computer users, damage computer systems, or encrypt documents for ransom, among other nefarious goals. In Q1 of 2021 alone, around 87.6 million new types of malware and 2.51 million new types of ransomware were detected, summing the total number of malware detected till 2021 to more than 1.51 billion and these figures keep growing constantly \citep{1}.

A prevalent way used in commercial antivirus products is using signature-based malware detection with signatures extracted by expert analysts but it has a small room for variation and is susceptible to evasion by obfuscation \citep{obfuscation}. Many machine learning-based malware analysis methods have been proposed \citep{schultz2001data} \citep{kolter2004learning} \citep{dai2009efficient} \citep{22} that automatically derive features from malware executables that are generalizable enough to counter current obfuscation techniques and can extend to new types of malware.

These machine learning-based approaches work by deriving static features to categorize malware. However, focusing only on static features may not represent the full semantic meaning of an executable \citep{static}. Deep learning-based approaches that can automatically learn representational feature space mappings from malware executable code have been proposed in an effort to have better generalizability \citep{17} \citep{19} \citep{3}. In recent years, the advancement in deep learning has enabled it to provide performance at par with what humans can do on several tasks \citep{silver2017mastering} resulting in growing faith in such real world deployed systems \citep{Tesla:2020} \citep{FaceID:2020} \citep{grigorescu2020survey}. However, deep learning systems are found to be vulnerable to adversarial attacks \citep{10}, which are malicious inputs specially designed to confuse a trained model to wrongly classify the output.

\section{Related Work}

Rule-based signature-based approaches require a cybersecurity researcher to manually set up rules, or categorize a binary as malware and mark its signature. This would require researchers to know how every new malware works and is not a scalable approach. \citep{17} propose a deep learning based approach to help solve this problem. \citep{18} describe using deep learning for malware detection as a double-edged sword, where deep learning could be really helpful in identifying new, yet unknown malware, but miscreants can also come up with ways to fool the neural networks by creating adversarial samples with small perturbations that do not change the sample’s original function, but rather fools the network into classifying it into some other class.

\citep{19} used CNNs to classify binaries as malware or benign files where binaries converted to an image representation were used. The authors were able to achieve best accuracy of 98.52\% for the Malimg dataset \citep{mailimg}, and best accuracy of 98.99\% for the Microsoft Malware Dataset \citep{mmd}. \citep{20} evaluated various methods of conducting adversarial attacks on CNN based malware detectors.
The success rate of white-box attacks for the Fast Gradient Sign Method (FGSM) was really low around 3\%, whereas for the Bit-Flip Attack (BFA) it was around a mean of 20\%.

After the success which recurrent neural networks have shown for other tasks, they have been tried for the task of malware detection \citep{1}. \citep{3} used a combination of convolutional neural networks and recurrent neural networks for the purpose of malware detection. RNNs were used for feature extraction and CNNs were used for feature classification. They obtain a best case AUC score of 0.96. With the use of RNN for malware detection, it became known that even they are susceptible to adversarial samples due to the general susceptibility of neural networks to adversarial attacks \citep{4}. To simulate the more realistic black-box nature of attacks, \citep{4} first trained a substitute RNN to simulate the behavior of the detector to be attacked. Another RNN was trained to create adversarial samples from malware inputs.

Previous methods did not look at the whole meaning of the assembly code, but rather looked at different chunks of the assembly language instructions. To overcome this, Transformer-based neural networks for malware detection were proposed by \citep{6}.
These Transformer-based approaches achieve better accuracy than previous approaches (\citep{21}, \citep{22}, \citep{17}, \citep{23}) in all experiments.

\section{Training a Transformer for malware detection} \label{training}

In this section we list down the details of training a competitive Transformers-based malware detector on which we will carry out an adversarial attack in section \ref{attack}, and evaluate defenses against the attack in section \ref{defense}.

\subsection{System Design and Architecture}
\begin{figure*}
    \centering
    \includegraphics[scale=0.25]{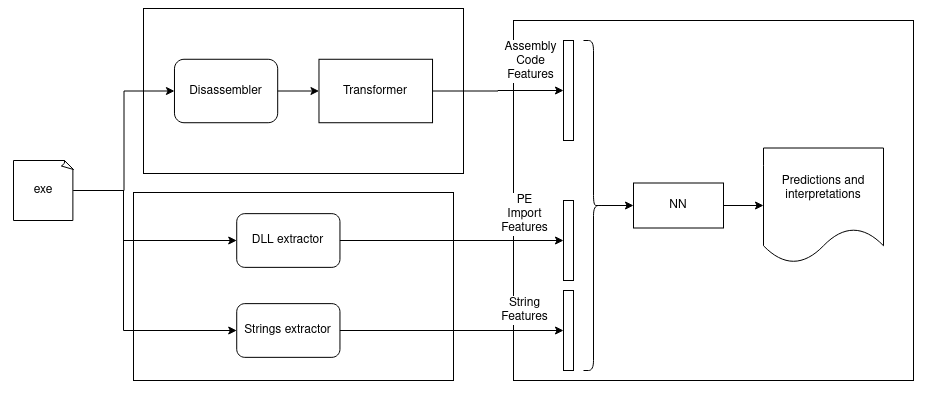}
    \caption{System Architecture}
    \label{fig:my_label}
\end{figure*}
Our malware detection system is mainly divided into 3 parts: 1. \textbf{Assembly Module} - The assembly module consists of a disassembler, a tokenizer and a Transformer. The input to the assembly module is an exe file, which is fed directly to the disassembler. The assembly module is responsible to calculate assembly language features, which would be used for final classification. 2. \textbf{Static Feature Module} - The static feature module consists of a DLL extractor, and a string extractor. The input to this is the same as that to the assembly module, an exe file. The DLL extractor extracts PE imports from the file, and the string extractor extracts all the printable strings from the given input file. The static feature module outputs two set of vectors, one from the DLL extractor, and the other from string extractor. The output from the static feature module will be used for final classification. 3. \textbf{Neural Network Module} - The neural network module consists of a neural network, which takes in assembly language features from the assembly module, and PE import features and string features from the static feature module, and performs a binary classification on whether the file is malicious or benign.

\subsection{Features}
The feed-forward NN also takes input static features extracted from the PE binaries.
The static features include 1. \textbf{PE Imports : }PE Imports are the DLL files imported by the given PE binary. This helps to capture the external function calls, and imports, and hence helps in categorizing suspicious files based on import patterns from the existing malware binaries \citep{17}. 2. \textbf{Printable strings : }All printable strings (only ASCII characters) of size greater than or equal to 6 are extracted from the given binary and used as another feature to train the vanilla NN.

The final feature set is thus formed by combining 1. the feature vector obtained from the Transformer, 2. the DLL feature vector obtained from Static Feature Extractor, and 3. the string feature vector obtained from Static Feature Extractor. The final feature vector is then fed to the fully connected neural network for classifying files as malicious or benign. 

\subsection{Dataset and Results}
We performed our expermients by collecting a total of 2985 malware samples from VirusTotal and 2215 benign executables from a fresh Windows 10 installation.
We used objdump to disassemble binary executables into .asm files. and tokenized strings to create a vocabulary which was eventually fed to the transformer. We were able to classify malware and benign files with a test set accuracy of 92.5\%.

\section{Adversarial attacks}

Machine learning models are vulnerable to misclassification when provided with a set of inputs that come from a different distribution than they were trained on. These inputs can be modified with special techniques to force misclassification. Such generated inputs are known as adversarial examples.

Notable examples of demonstrated adversarial attacks include McAfee’s fooling of Tesla autonomous vehicle by just adding 2 black strips on a speed limit sign making the Tesla AV go 50 miles per hour past the limit \citep{8}. Dresswear that can fool face detection systems or license plates that can fool automatic license number capture systems use adversarial examples to make such models misclassify \citep{9}.

Adversarial attacks can be classified into two broad types of Evasion attacks and Poisoning attacks. An evasion attack is used when the attacker does not have access to the model during the training phase. In this work, we consider evasion attacks as we consider that to be a generalized case where an attacker does not have access to the training of the model. Evasion attacks can be broadly divided into the two steps of estimation of sensitivity for each direction of perturbation and selection of directions for perturbation. 

Some of the algorithms used for the estimation of sensitivity are - 
\begin{itemize}
    \item \textbf{L-BFGS - }Introduced by \citep{10}, this method tries to solve the minimization problem of finding the minimum perturbation that can force misclassification with the L-BFGS optimization method.
    \begin{equation}
        arg\;min\;_r\;f(x+r) = l\qquad(x + r) \in D
    \end{equation}
    where \(l\) is not equal to the target label \(h(x)\).
    \item \textbf{FGSM - } \citep{11} introduced the Fast Gradient Sign Method that can solve the equation above computationally efficiently.
    \begin{equation} \label{fgsm}
        X^* = X + \varepsilon * sign(\nabla_x C(X, Y_{true})
    \end{equation}
    Here \(C\) is the cost function used in the model. \(X^*\) is the adversarial counterpart of the input \(X\). \(\varepsilon\) is the amount of perturbation. \(\nabla_x\) is the gradient of the cost function.
    \item \textbf{Jaccobian Method - } \citep{12} determined sensitivity in each dimension by finding out Jacobian of the trained model, i.e. it’s derivative in a forward way and then perturbed input in most sensitive dimensions.
\end{itemize}

\section{Attacking our trained Transformer} \label{attack}

We used the Fast Gradient Sign Method by \citep{11} to craft out adversarial samples to fool the Transformer-based malware detector trained in section \ref{training}. FGSM can be implemented in two ways, either by using the target class directly or by using the iterative method. We used the target class directly to generate our adversarial samples by substituting $Y_{true}$ with $Y_{target}$ in equation \ref{fgsm} above.
\begin{equation}
    X^* = X + \varepsilon * sign(\nabla_x C(X, Y_{target})
\end{equation}
We perturbed all possible dimensions and achieved a misclassification rate of 23.9\% with the Fast Gradient Sign Method.

\section{Defenses against adversarial attacks} \label{defense}

With the increase in adversarial attacks on commercial systems, many defenses have been proposed against them. Although these defenses do not provide complete immunity against adversarial samples, they act as deterrents. Some defenses are listed below.
\begin{itemize}
    \item \textbf{Training on adversarial samples - } This is a brute force approach where the input distribution of the model is expanded. Although this defense is not very useful in the case of black-box attacks as shown by \citep{13}, it is still widely used as a practical defensive approach against adversarial samples.
    \item \textbf{Masking the gradient of the model - }Many attack methods including FGSM depend on the derivative of the trained model. Nearest neighbor classifiers or decision trees-like models can effectively deter such an adversarial attack. However, these methods often underperform when compared to neural architectural methods. 
    \item \textbf{Reducing feature space - }Developed by \citep{14}, this method reduces the number of features and with it the number of dimensions to add perturbations to declines naturally.
    \item \textbf{Transferability block - }Adversarial attacks are successful in many cases due to the transferability property of neural networks. \citep{15} block transferability by training models to output NULL to peturbed inputs.
\end{itemize}

In our approach, we set up defenses for our model with two defenses from the ones listed above. With the most practical adversarial training, the misclassification rate dropped to 11.2\%. With reducing the feature space, we did not get promising results, and the misclassification rate reduced by a mere 2.4\% to 21.5\%.

\section{Conclusion and Future Scope}
The use of deep learning techniques for the task of malware detection has given promising results and is in use at a few of the most sought after anti-malware products \citep{16}. But due to the inherent nature of such deep learning techniques, these malware detectors are prone to adversarial attacks. We have implemented an avant-garde machine learning detector using Google’s Transformer neural network architecture, demonstrated an adversarial attack on the same, and proposed defenses against such adversarial attacks. The future scope in this directoin could aim at demonstrating more types of adversarial attacks on such malware detectors and propose better defenses that do not need access to the trained model.

{
\small
\bibliography{main}
\bibliographystyle{plainnat}
}

\end{document}